# Polar nature of the ferroelectric nematic studied by dielectric spectroscopy

Neelam Yadav[1], Yuri P. Panarin*[1,2], Jagdish K. Vij[1], Wanhe Jiang[3], Georg H. Mehl*[3]

[1]Department of Electronic and Electrical Engineering, Trinity College, Dublin 2, Ireland
[2]Department of Electrical and Electronic Engineering, TU Dublin, Dublin 7, Ireland
[3]Department of Chemistry, University of Hull, Hull HU6 7RX, UK

**Abstract**: The nematic-nematic transitions in a nematic compound DIO are studied in homogeneously planar and homeotropic aligned cells using dielectric spectroscopy in the frequency range 0.1 Hz to 10 MHz over a wide range of temperatures. Three relaxation processes are identified in this material. All the three relaxation processes show large jumps in the dielectric strength and discontinuity in frequency at the $N_X - N_F$ transition, indicative of it being a first order phase transition. Unlike a conventional nematic liquid crystalline material that usually shows molecular relaxation processes at higher frequencies, this material shows three processes at relatively lower frequencies. The three processes are found to be collective in nature. The least frequency process is proven to be due to the conductivity of ions in the medium and due to the accumulation of charge at the alignment layers unlike mistakenly reported in the literature otherwise. The mode at the highest frequency is proven to be due to the ferroelectric domains in the $N_F$ phase. This is evidenced by its dielectric strength two orders of magnitude higher in a homeotropic aligned cell than planar aligned cell. Its mechanism is soft mode like in the N phase. The intermediate frequency mode is proven to be due to the correlation of molecules of the medium as is normally observed in liquid crystalline cybotactic clusters.

Yuri.panarin@tudublin.ie
g.h.mehl@hull.ac.uk

**Introduction:** Since the discovery of liquid crystalline state in the various derivatives of cholesterol, the chirality and polarization properties of liquid crystals have only been associated with molecular chirality where the carbon atom has four different substitutions. In the last couple of decades, such features have also been observed in other LC systems built from non-chiral molecules, such as the bent-core and bent bi-mesogen LCs. In such systems achiral molecules may form so-called "chiral phases" which possess optical activity and/or spontaneous polarization. Recently, a new, long-awaited Ferroelectric Nematic phase predicted by P. Debye [1] and M. Born [2] more than a century ago, was finally observed. In the first publication [3] related to ferroelectric nematic, extremely high dielectric permittivity $\sim 10^4$ and huge spontaneous polarization $\sim 4.4$ μC cm$^{-2}$ in a material called DIO was reported. This has been attributed to the hysteresis like switching behavior under external electric field, typical of ferroelectric LCs. The material, DIO, exhibits several modifications of nematic phases, such as the ordinary paraelectric nematic (M1 or N), high polarizability nematic (M2 or Nx or SmZ$_A$) and ferroelectric nematic (MP – or N$_F$) [4,5]. Independently, Mandle et al synthesized a range of compounds with large dipole moments RM230, RM734 and their homologues which display two distinct nematic mesophases, named as N and N$_X$, separated by a weak first-order transition [6, 7, 8, 9]. The polar N$_X$ phase formed the so-called splay nematic phase (N$_S$) [10,11,12] due to strongly non-calamitic molecular shape with the splay modulation period of 5–10 microns and the N-N$_s$ phase transition bears resemblance to the ferroelectric to ferro-elastic transition which takes place via flexoelectric coupling [12]. Chen et. al. gave the first principle demonstration of ferroelectricity at lower temperatures in RM734 and termed the previously known N$_X$ or N$_S$ phase as ferroelectric nematic (N$_F$) phase having striking different electrooptical properties [13]. The N$_F$ has been observed in a bent core material BC-F as well [14]. The material DIO was revisited by the Boulder group in 2021 [15]. They found that the intermediate (MP) phase that exists between the paraelectric, uniaxial, homogeneous N and the ferroelectric, uniaxial, unmodulated N$_F$ phase is biaxial, density modulated with period of 8.6 nm, splay modulated and antiferroelectric. They have designated this phase as SmZ$_A$ phase. These previously unknown phases display fascinating properties and hence are being studied by several research groups around the world in order to understand the phenomenon better [16]. Recently, Manabe et al. synthesized a compound that shows a direct transition from the isotropic to N$_F$ phase [17] while Saha et al. reported a polar LC material that exhibits multiple ferroelectric nematic phases [18].

It is important to understand the novel nematic-nematic phase transitions with a view to harness enormous potential for their applications in supercapacitors, sensors and photonics. The origin, structure, and the mechanisms for the formation of these multiple nematic phases are not yet fully understood and are currently being vigorously debated. This lack of understanding and the scientific know-how is hampering the development of technological applications. Hence, the characteristics of the N, N$_x$ or SmZ$_A$ and N$_F$ phases are being investigated using the technique of dielectric spectroscopy in planar and homeotropic aligned cells in an achiral rigid linear LC (called DIO) with the phase sequence Iso - N – N$_x$ - N$_F$.

**Results and Discussion**

Dielectric spectroscopy is a powerful method for studying the relaxation phenomena in materials. While it doesn't supply direct information about the micro/nano structure of material, nevertheless these can be indirectly deduced from the dielectric spectra. The technique complements results obtained by methods such as X-ray diffraction, SHG, and different types of microscopies. Dielectric spectroscopy involves the response of electric dipoles to the weak external electric field and the technique is an excellent probe in detecting changes in the structure of the medium brought about its constituent molecular dipoles. Also, dielectric studies are superior to the electro-optical methods for observing the relaxation modes present in the LCs as during measurements, the structure of the phase is not significantly perturbed by the field. The anisotropic nematic interactions influence the structural distributions favoring molecular geometries acclimatizing nematic ordering. The orientational ordering of LC phase and the rotational distribution of dipoles in the presence of a weak probe field determines the temperature dependent dielectric response of the LC phase. Dielectric spectroscopy was applied for investigation of ferroelectric nematics and in a number of publications [19, 20, 21] including the DIO material [3**Error! Bookmark not defined.**, 21]

We have successfully employed this in experimental study of different phases and materials such as ferroelectric SmC [22], TGBA [23], ferrielectric [24, 25] and antiferroelectric SmC-like LCs [26], de Vries phase [27], bent-core nematics [28], SmAP [29], SmCP [30].

In this paper, we investigate the ferroelectric nematic material, DIO [3] by dielectric spectroscopy. The molecular structure and the phase sequence of DIO as presented in Ref. [15**Error! Bookmark not defined.**] is shown in Fig.1,

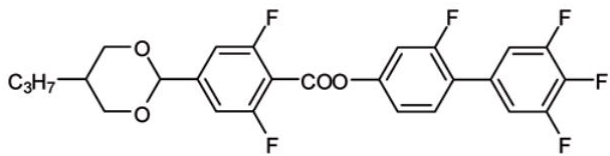

On cooling: Cr 34 MP 68.8 M2 74.5 M1 190 Iso (˚C)

On heating: Cr 68 M2 84 M1 190 Iso (˚C)

**Figure 1.** (a) Molecular structure and the phase sequence of DIO [3].

The dielectric permittivity measurements are carried out using a broadband Alpha high-resolution dielectric analyzer (Novocontrol GmbH, Germany). The sample cells for dielectric measurements are constructed from glass substrates coated with indium tin oxide (ITO) with low sheet resistance (10-20 $\Omega/\square$). This shifts the parasitic dielectric peak to a higher frequency due to the capacitance of the cell in series with the finite resistance of the ITO. The capacitance of the empty cell was also measured. These measurements on the aligned liquid crystalline sample were carried out under cooling from 190 to 40 °C. The temperature was varied in steps of 1 °C under the application of a weak voltage of 0.1 V. Temperature of the cells with the sample aligned was stabilized to within a range of ±0.02 °C

The temperature dependence of the real and imaginary components of the complex permittivity in the frequency range of 0.1 Hz to 10 MHz were measured in both homeotropic and planarly aligned cells of 4 μm thickness under slow cooling from the isotropic phase. The temperature dependent dielectric loss spectra of longitudinal ($\varepsilon_\parallel''$) and transverse ($\varepsilon_\perp''$) components of 4 μm thickness cells are shown in the Figure 2. Figure 3 presents dielectric loss in four different phases: Isotropic 190 °C; M1 (N) 150 °C, 130 °C, 90 °C; M2 ($N_X$) 70 °C; and MP ($N_F$) 90 °C. At first glance on Figs. 2-3, several unusual features of these spectra are noticed.

- Unlike the ordinary liquid crystalline nematic which shows only the molecular relaxation processes, there are up to three collective relaxation processes in the nematic phase of DIO. These are termed as P0, P1, and P2 with increasing relaxation frequency.
- The low-frequency conductivity in a homeotropic aligned cell is much higher than in a planar homogeneous aligned cell.
- The strength of the conductivity mode in the homeotropic cell initially decreases on cooling but at lower temperatures it starts to increase.

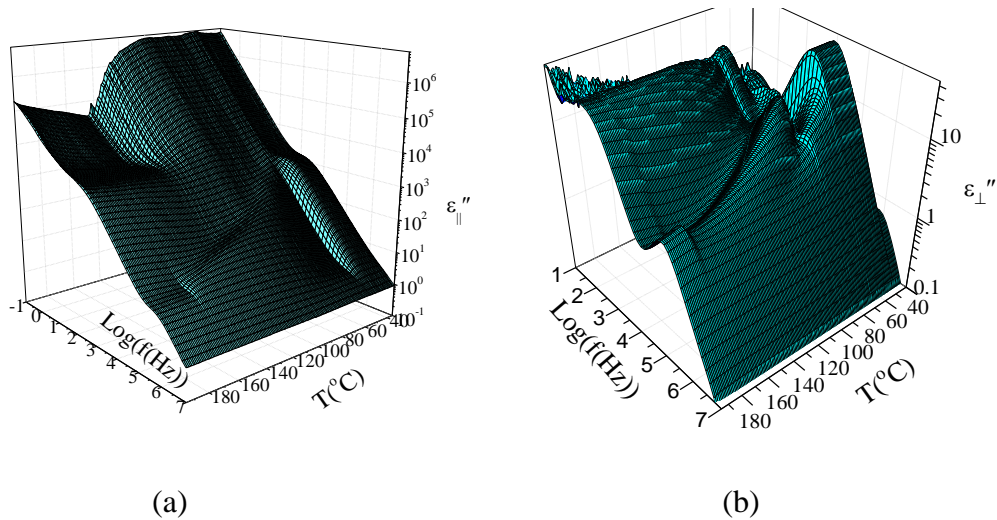

(a)                                                    (b)

**Figure 2**. 3D Dielectric loss spectra of DIO in 4 μm (a) homeotropic and (b) planar homogeneously aligned cells.

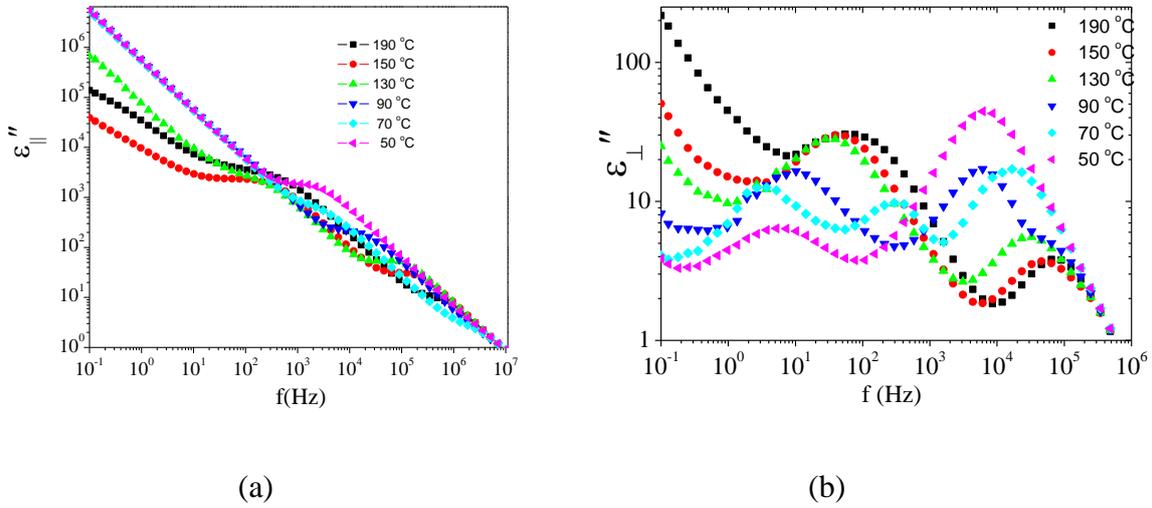

(a)                                          (b)

**Figure 3**. 2D cuts of 3D dielectric loss spectra (Fig.2) in four different phases: Isotropic 190 °C; M1 (N) 150 °C, 130 °C, 90 °C; M2 (N$_X$) 70 °C; and MP (N$_F$) 90 °C.

To clarify these unusual features and to determine the physical origin of the relaxation processes, the dielectric spectra of the sample cells are analyzed using WINFIT software of Novocontrol GmbH. The dielectric data on the complex permittivity are fitted to the Havriliak - Negami equation (Eqn. (1)).

$$\varepsilon^*(\omega) = \varepsilon' - i\varepsilon'' = \varepsilon_\infty + \sum_{j=1}^{n} \frac{\delta\varepsilon_j}{\left[1 + (i\omega\tau_j)^{\alpha_j}\right]^{\beta_j}} - \frac{i\sigma}{\varepsilon_0\omega} \qquad (1)$$

where $\varepsilon_\infty$ is the high frequency dielectric permittivity depending on the electronic and atomic polarizabilities of the material, $\omega = 2\pi f$ is the angular frequency of the probe field, $\varepsilon_0$ is the permittivity of free space, $\sigma$ is dc conductivity, $\tau_j$ is the relaxation time of the j$^{th}$ process, $\delta\varepsilon_j$ is the dielectric amplitude or strength of relaxation process, $\alpha_j$ and $\beta_j$ are the symmetric and asymmetric broadening parameters that determine the distribution of the relaxation times of the j$^{th}$ process. In general, the data here were fitted to three processes, i. e. $n = 3$. For the relaxation processes P1, P2 the fitting parameters were $\alpha_j = 1$ and $\beta_j = 1$ which correspond to the Debye type relaxation, while for the low-frequency relaxation processes P0, $\alpha_j < 1$ and $\beta_j = 1$ being of the Cole-Cole type. The temperature dependence of the relaxation processes in terms of the dielectric strength and the relaxation frequency in homeotropic and planar homogeneously aligned cells are shown in the Figure 4.

### *Properties and physical nature of the relaxation process P0.*

The symmetrical stretching parameter of relaxation processes P0 $\alpha_0 < 1$ being of the Cole-Cole type, this mode depends strongly on the LC cell parameters, especially the thickness of the alignment layer. We identify this process as the ion separation process since it also exists in the ordinary nematics, e.g. 5CB, where no collective relaxation processes are observed However, the practical LC cells incorporate alignment layers for achieving certain defined orientations, homogeneous or homeotropic. The alignment layer limits conductivity, i.e. on application of the external electric field, the ions initially distributed uniformly in the volume of the cell start moving in the direction of electric field and accumulate on the surface of the alignment layers. On applying an AC electric field, these respond to the alterations of the field and thus produce a relaxation process in the dielectric spectra. Its relaxation frequency depends on the ion mobility (and therefore on temperature) and thickness of the cell. The dielectric strength of this process also depends on ion mobility, their concentration and the electric conductivity of the alignment layers which depends on the thickness and the nature of the alignment layers. There are two opposite scenarios: (i) finite conductivity and (ii) zero conductivity. In the first case we would expect only a low-frequency exponential conductivity term ($\frac{i\sigma}{\varepsilon_0\omega}$) in Eq.(1). In the opposite case there should be no conductivity term, but a strong relaxation process will arise from ions separation and their accumulation on layers.

Practically, however, both terms exist in the dielectric spectra. We have studied the dielectric spectra of 5CB in three different cells: (i) 1.6 μm with SiO protecting dielectric layer; 2 μm commercial planar cell; and hand-made 2 μm cells without alignment layer. In the 1.6 cell for 5CB with a SiO dielectric layer, the conductivity term is much smaller than the dielectric strength P0 and the opposite occurs for the cell without an alignment layer. In typical LC cells, commercial and home-made, with the alignment layer thickness of ~100 nm, both terms are distinguishable. Unfortunately, occasionally this parasitic process is treated in the literature as the collective relaxation process which in turn leads to the incorrect conclusions with regards to the occurrence of the spontaneous polarization.

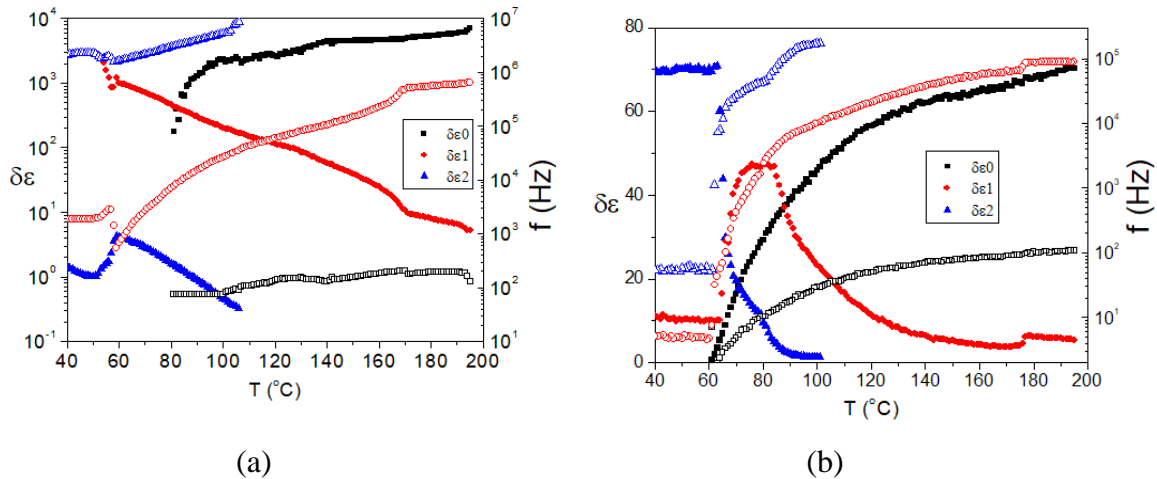

(a)                                    (b)

**Figure 4.** Temperature dependence of relaxation process: dielectric strength (δε) (filled) and relaxation frequency(f) (unfilled) in 4 μm (a) homeotropic and (b) planar homogeneously aligned cells.

### _Properties and the physical nature of the relaxation process:_ P1

The temperature dependence of dielectric parameters of the mid-frequency relaxation process P1 are shown in the Fig.4 (in red). This relaxation process exists in the temperature range 30-200 °C including in the isotropic phase. Hence we can confidently assign it to the molecular relaxation process around the short axis. In the isotropic phase $\varepsilon_{iso} = (\varepsilon_{\parallel} + 2\varepsilon_{\perp})/3$, therefore on transition from the isotropic state (Iso) to the N phase the dielectric strength of this process should jump up to $\varepsilon_{\parallel}$ in a homeotropic cell and jump down to $\varepsilon_{\perp}$ in a planar cell and this is in good agreement with Fig.4. This finding supports our assignment that the process P1 is a relaxation process around the short axis.

However, the temperature dependence of the dielectric parameters ($\delta\varepsilon$, f) has certain peculiarities, the temperature dependence of the relaxation frequency of the molecular relaxation process follows an Arrhenius law $f(t) = Ae^{\frac{-E_a}{RT}}$, i. e. it should look like straight line if plotted on a log Y-axis, which contradicts the fitting results for f2, as shown in Fig.4. Instead, the experimental temperature dependence shows soft mode - like behavior, i.e. the relaxation frequency (and the inverse dielectric strength) decrease linearly on approaching the phase transition from $N_x$ to $N_F$ phases, $T_{NXF}=$ 67 °C as $\sim$(T-$T_{NXF}$) [31] in both homeotropic and homogeneous cells. This unusual behavior reminds us of similar results in the nematic phase formed by bent-core molecules [32] which were explained due to the formation and growth of so-called cybotactic clusters. In the isotropic phase of DIO molecular relaxation occurs with individual molecules. On the transition to the nematic phase the relaxation becomes correlated with neighboring molecules at correlated distances (length) which leads the relaxation frequency to decrease jump like (see Fig.4). On further cooling the correlation length (and volume) continues to grow on cooling to the $N_x$ to $N_F$ phase transition, showing soft mode - like behavior with signal saturation in the $N_F$ phase. This mode is a high polarizability mode.

### _Properties and physical nature of relaxation process P2._

In general, the temperature dependence of the dielectric parameters of the high–frequency relaxation process P2 are similar to those of the mid-frequency relaxation process P1, with some differences:

- The process does not exist in the isotropic phase but appears in the nematic phase at ~110 °C. Actually, there is nothing specific happening at this temperature and P2 exists even at higher temperatures but is screened by other two strong modes and at this temperature it just becomes distinguishable.

- the dielectric strength of this mode increases monotonically on cooling to the $N_x$ to $N_F$ phases transition and then it jumps up in value and saturates in the $N_F$ phase in homeotropic cells but jumps down and saturates in planar cells.

Combining the two experimental findings that: (i) P2 does not exist in the isotropic phase and (ii) it appears in the nematic phase that gradually transforms to a ferroelectric mode we can conclude that this relaxation process is linked to the formation of the ferroelectric domains, which grow on cooling and occupy finally the entire volume in $N_F$ phase.

Finally, Figure 5 shows the dielectric anisotropy which is positive over the entire temperature range investigated in this study (up to 195 ° C).

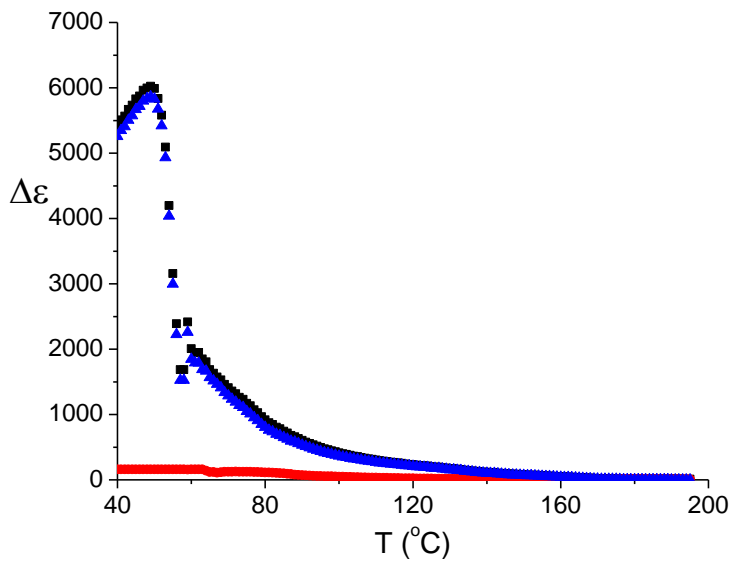

**Figure 5**. Temperature dependence of dielectric anisotropy $\Delta\varepsilon = \varepsilon_\parallel - \varepsilon_\perp$. There is a discontinuity at the N-NF transition temperature.

### *Conclusion.*

In the present manuscript we studied ferroelectric nematic material, DIO [3], by dielectric spectroscopy in the frequency range of 0.1 Hz to 10 MHz and in the temperature range from 195 ˚C to 35 ˚C. In total three collective relaxation processes were observed, where the least-frequency relaxation process P0 is due to ion separation of ionic impurities, P1 is a soft-mode like polarizability process of correlated molecules volume, and P2 is due to the appearance and growth of ferroelectric domains.